\documentstyle[psfig,epsf,epsfig]{europhys}

\def\And{{\rm and\ }}

\newif\ifboo \boofalse

\begin{document}
%
%
%
\euro{xx}{x}{xxx}{2000}
\Date{}
\shorttitle{A. Barrat and V. Loreto, Memory in aged granular media}
%
%
%
\title{Memory in aged granular media}
\author{A. Barrat\inst{1} \And  V. Loreto\inst{2} }
\institute{
\inst{1} Laboratoire de Physique Th{\'e}orique,
Unit{\'e} Mixte de Recherche UMR 8627,\\ 
B{\^a}timent 210, Universit{\'e}  de Paris-Sud 
91405 Orsay Cedex, France; \\ e-mail: Alain.Barrat@th.u-psud.fr\\
\inst{2} Universit\`a degli Studi di Roma ``La Sapienza'',
Dipartimento di Fisica, \\ P.le A. Moro 5, 00185 Rome, Italy
and INFM, Unit\`a di Roma 1;\\
e-mail: loreto@pil.phys.uniroma1.it
}
%
%
%
%
\pacs{
\Pacs{05.70.Ln},{81.05.Rm},{75.10.Nr}
      }
\maketitle
%
%
%
\begin{abstract}
 Stimulated by recent experimental results, we simulate
``temperature''-cycling experiments in a model for the compaction of granular
media. We report on the existence of two types of memory effects:
short-term dependence on the history of the sample, and long-term
memory for highly compact (aged) systems. A natural interpretation of these
results is provided by the analysis of the density heterogeneities.
\end{abstract}
%
%
%
%
%
%
The study of the rheology of granular media represents nowadays an important
chapter in the more general context of the so-called soft condensed
matter \cite{grain}.
The general questions one is interested in concern the response behaviour
of a generic granular medium submitted to an external perturbation. In this
spirit it has been recognized that the response of such systems 
depends in a non-trivial way on the mechanical properties of the grains 
composing the medium, on the  boundary conditions, on the driving procedure
(the way to inject energy to perturb the system) and, last but not the least,
on the past history of the system, e.g. on the procedures the 
system has undergone until the moment we study it. 

In this context the similarities between granular media and
the phenomenology of glassy systems have been exploited.
One of the main results has been to recognize that in  a compaction
procedure, where a certain system increases its density if shaken
or tapped, a granular medium displays ageing \cite{struik} and its 
properties, e.g. the two-times correlation function, depend on the
time where one performs the measurement\cite{nicodemi,bl}, 
i.e. on the age of the system.

The analogy with glassy systems has mainly allowed 
to export tools and techniques from one field to the other.
Typical experiments on glassy systems consists in monitoring 
the effect of changes imposed to the external temperature on the
macroscopic observables \cite{SG}. Translated in the framework of granular 
matter, i.e. of a non-thermal system, this means to analyse the effect
of changes in the driving procedure, e.g. in the tapping 
amplitude $\Gamma$, which, in a typical compaction experiment, 
represents the control-parameter. 

In a previous work \cite{bl}, we have presented a detailed study, in the 
framework of a microscopic model for granular media,
focusing on the response to small
perturbations in the driving rate.
We have pointed out the importance of the interplay 
between the response properties and the spatial structures
that spontaneously emerge as a consequence of the dynamics imposed 
to the system. In particular we have studied the relevance of the 
densities heterogeneities in the process of granular compaction,
and shown that the density is not the only relevant
information about the system. Configurations with the same density
but reached with different dynamical procedure show completely 
different rheological properties. This was nicely confirmed
by a set of recent experiments\cite{joss}: three systems prepared 
at the same density but in different ways display different behaviours 
if the same tapping acceleration is applied to them. 
This feature concerns the coding of the system history in the microscopic 
configurations.

In the same set of experiments, similar in spirit to the temperature-cycling 
experiments in spin-glasses\cite{SG} the authors of\cite{joss}
have put in evidence 
the existence of memory effects in the compaction of granular media.
In particular they performed the study of the effect of relatively 
large and abrupt changes in the tapping acceleration. 
This is an efficient way to obtain reliable experimental data
since the effect of small changes in the tapping acceleration
is difficult to detect in real experiments.
What was shown in \cite{joss} is that, during the compaction, the response
to an abrupt change in the tapping acceleration $\Gamma$ 
is opposite (at least at short times) to what could be expected
from the long-time behaviour of the compaction:
while at constant $\Gamma$ a higher compaction rate is obtained for
larger $\Gamma$, a sudden increase in $\Gamma$ leads to a decompaction,
and a sudden decrease gives rise to a temporary increase in the
compaction rate. This behaviour is however transient, and the usual
compaction rate is recovered after a while: the authors therefore
call this effect ``short-term memory''. The same qualitative behaviour
has been obtained also in experiments of compaction under shear 
\cite{pouliquen}.

In order to push forward this kind of studies we have extended our work
\cite{bl} to study the response to large changes in the driving.
In particular we explore the behaviour of the Tetris Model
\cite{tetris,RTM} for abrupt changes in the driving rates in various 
time regimes, 
that were not considered in the preliminary, time-limited, 
experiments~\cite{joss}.  We also explore the return to the initial
tapping acceleration and the presence or absence of memory, in a
way similar to the temperature-cycling experiments in spin-glasses.
We perform these temperature-cycling experiments in non-equilibrium
(aging) as well as in quasi-stationary (time-translation invariant)
situations.   
We hope in this way to stimulate new and more precise experiments along
the same lines.

The model we consider, the so-called Tetris Model\cite{tetris},
consists of particles on a tilted (two-dimensional) square lattice,
which has lateral periodic boundary conditions
\footnote{the use of closed lateral boundary conditions
do not change the results} and a closed boundary
at the bottom. With respect to its original definition we consider
a random version where each particle can be schematized in general
as a cross with  $4$ arms (in general the number of arms is equals
to the coordination number of the lattice) of different lengths,
denoted by $l_{NE}$, $l_{NW}$, $l_{SE}$, $l_{SW}$,
chosen in a random way (see \cite{RTM,bl} for details).

Gravity and shaking are implemented as follows: the particles
diffuse on the lattice, with geometrical constraints, and
probability $p_{up}$ to move upwards (with $0 < p_{up} < 1$),
and a probability $p_{down}=1-p_{up}$ to move downwards.
The quantity  $x=p_{up}/p_{down}$ can be related 
to the adimensional acceleration
$\Gamma$ used in compaction experiments\cite{exp-compaction} through the 
relation $\Gamma \simeq 1 / log(1/\sqrt x)$. The sizes (width
$\times$ height) used ranged
from $200 \times 60$ to $20,\ 40,\ 60 \times 200$,
 in order to test various aspect ratios.

\begin{figure}[7]
\centerline{
       \psfig{figure=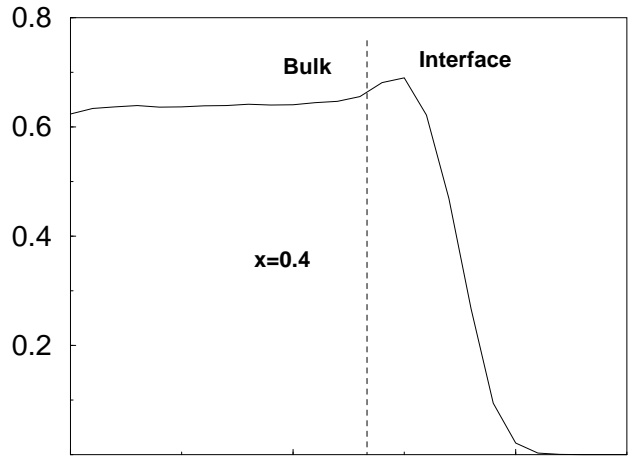,width=6cm,angle=-90}
}
\caption{Typical density profile obtained after $t=10^4$ Monte-Carlo
steps at constant forcing $x=0.4$.}
\label{fig:profile}
\end{figure}

The procedure used in the experiments and in our simulations
is inspired by classic experiments in spin-glasses. 
The system is initialized by random addition of particles, 
and let evolve with a constant external forcing $x_1$
up to a certain time $t_w$.
Two copies of the system are then made:
one copy is kept evolving with $x_1$ at all times for reference, while
the second copy evolves with a different forcing $x_2$
for a time $\Delta t$, after which the forcing is set
back to $x_1$.
We monitor the mean height,
the bulk density (defined as the average density in the lower
$50$ of the system), and the density profiles of both copies
during the whole evolution for various values of the difference $x_1-x_2$, 
$t_w$ and $\Delta t$. We recall that the density profiles are defined 
averaging the density along the horizontal direction, giving thus indications
on the heterogeneities along the vertical direction (see figure 
\ref{fig:profile}), and allowing to roughly distinguish two regions, the bulk
and the interface. Before analyzing
the different regimes, let us recall a few general features
of the compaction at constant forcing. At short times, the density
is an increasing function of the forcing while at long times, there exists 
an optimal value of $x$ that allows to obtain the maximal density in the
fraction of the system considered for the measurement. This phenomenology
has been shown to be linked to the presence of heterogeneities in the system
\cite{bl}.

\begin{figure}
\centerline{
       \psfig{figure=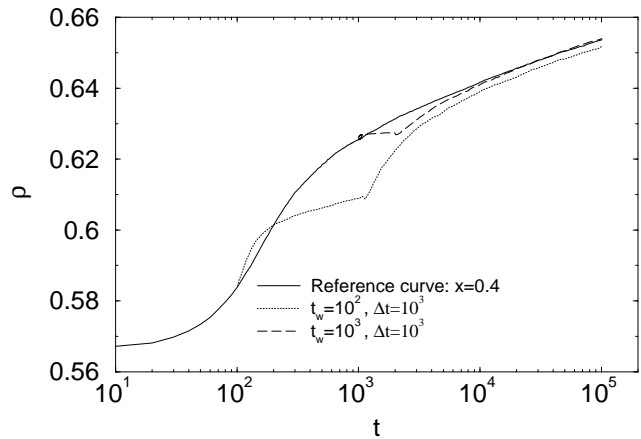,width=6cm,angle=-90}
}
\caption{Density versus time, for a reference case with constant $x=0.4$,
and with $x_1=0.4$, $x_2=0.1$, for various values of 
$t_w$ and $\Delta t$. 
The first effect at $t_w$ is an acceleration of the compaction,
but at larger times the curves cross; after $t_w+\Delta t$ the compaction
is again fast.
}
\label{fig:.4.1}
\end{figure}
\begin{figure}
\centerline{
       \psfig{figure=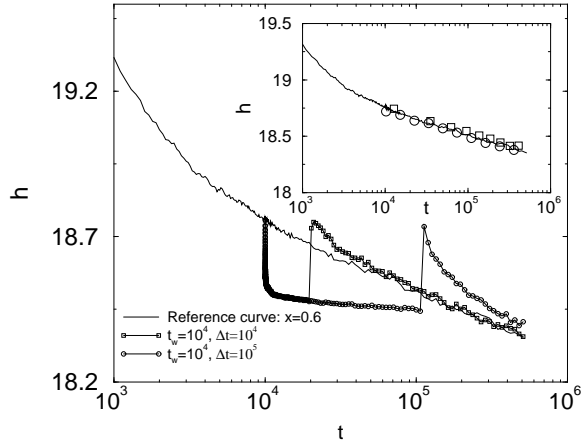,width=6cm,angle=-90}
}
\caption{Effect of a decrease of the forcing with $x_1=0.6$,
$x_2=0.3$, $t_w=10^4$, and $\Delta t=10^4$ and $10^5$:
at short times after $t_w$ the mean height $h$
decreases strongly (corresponding to an increase in the compaction, as in
figure 1), but remains then almost constant. As $x$ is again
increased, $h$ goes back to its value at $t_w$. The 
long time memory effect is illustrated in the inset:
the symbols, corresponding to the data for $\Delta t=10^4$ and 
$\Delta t=10^5$, are shifted by $\Delta t$ and they coincide with 
the reference curve.}
\label{fig:.4.1bis}
\end{figure}

{\bf First case: $\mathbf{x_1 > x_2}$.} 
When, at $t_w$, $x$ is suddenly lowered, the compaction rate of
the perturbed system first increases. This
behaviour, shown in figure \ref{fig:.4.1} and \ref{fig:.4.1bis}, 
is in agreement with the experimental results,
and it is opposed to the results one would
expect looking at the compaction data at constant forcing. This shows
that the system has some memory of its history at $t_w$. 
After a transient, however, this memory is lost:
compaction slows down, and the rate of compaction 
crosses over to the one observed at constant forcing:
the curves of the reference and perturbed systems therefore cross. 

While the transient, corresponding to the short-term
memory, is observed for all values of the parameters investigated,
an important difference appears for different $t_w$ at long times: if 
$t_w$ is small, the system is not very compact,
and is able to evolve a lot during $\Delta t$. The system does not display
memory for times larger than $t_w+\Delta t$. On the other hand, if the system is already quite
compact at $t_w$ (i.e. if it is {\bf aged} enough), the system evolves
very slowly during $\Delta t$, and for $t \ge t_w+\Delta t$
the compaction curves can be translated and superimposed to the reference one,
as shown in the inset of figure \ref{fig:.4.1bis}. 
During $\Delta t$ the system is able to keep 
memory of its state at $t_w$. 
This shows that a sufficiently aged system displays a long term memory, in
addition to the short-term memory existing at all times.



The interpretation of these results is quite straightforward using
the results of \cite{bl} and looking at the density profiles along the
vertical direction:
when $x$ is abruptly lowered, the first effect is that the particles
tend to go down, and the interface becomes steeper and more compact.
Therefore the density first increases with respect to the unperturbed case.
At larger times however, the evolution
is slowed down by the creation of a dense
layer at the interface, which blocks the bulk rearrangements needed
for the compaction \cite{bl}. 
After $t_w+\Delta t$, the increase in the forcing allows to
suppress the dense layer,
and the compaction can become again fast.
Moreover, if $t_w$ is large enough (typically $10^3$ or $10^4$ MC steps),
the bulk of the system is already quite compact,
and therefore the smaller value of the forcing 
during $\Delta t$ leads to a compaction of the
interface but the bulk almost does not evolve. At $t_w+\Delta t$, the 
forcing is again increased: the relaxation of the interface being fast, this 
leads the system back to its state at $t_w$. The inset of Figure 
\ref{fig:.4.1bis} illustrates
this behaviour: the curves
of compaction after $\Delta t$, shifted backwards by $\Delta t$, 
superimpose with the reference
curves. This is not the case for lower values of $t_w$,
for which the bulk is not yet highly compact at $t_w$, and 
evolves a lot even with a small driving $x_2$.

This memory effect, similar to what can be obtained in spin-glasses when
the temperature is first lowered then increased back to its previous value, 
was unfortunately
not investigated in \cite{joss}: it can indeed occur only at long times. It
would certainly be very interesting to have experimental checks
on this aspect.

\begin{figure}
\centerline{
       \psfig{figure=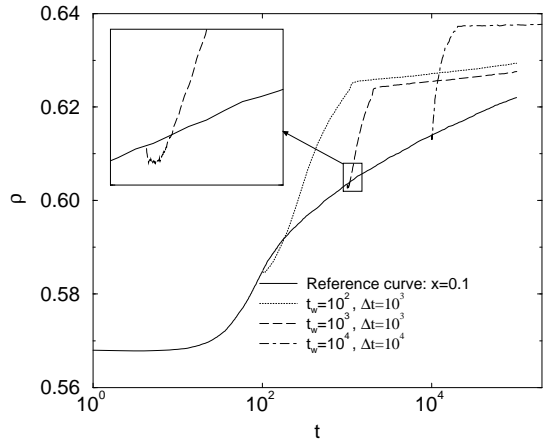,width=6cm,angle=-90}
}
\caption{Effect of an increase in $x$
(i.e. $x_1 < x_2$), for $x_1=0.1$, $x_2=0.4$, and various values of $t_w$
and $\Delta t$; the inset shows a zoom of the decompaction 
effect at short times
after the change in $x$. The compaction rate then is increased during 
$\Delta t$, and decreases again after $t_w+\Delta t$.}
\label{fig:.1.4}
\end{figure}

{\bf Second Case: $\mathbf{x_1 < x_2}$.}
When $x_1 < x_2$ we observe a short-term memory effect, 
as in the previous case. First, as the forcing is increased, 
one observes a decompaction; later on, the fact that $x$ 
is larger prevails and the compaction proceeds faster, at the normal
rate for constant $x=x_2$. The memory of the history up to $t_w$
is lost after a transient (see figure \ref{fig:.1.4}).
These features are again similar to the experimental
results. Moreover, at long times no memory is kept, since
the whole system evolves a lot during $\Delta t$.

Once again, the study of the density profile allows for an interpretation
of these results. As the tapping intensity is increased, the first effect
is a decompaction, especially at the interface. The fact that the interface
is less compact then allows for a much better compaction of the bulk. Note
that this behaviour is closely related to the existence of a response function 
that is positive at short times and negative at longer times \cite{bl}.
At $t=t_w+\Delta t$, the bulk has been deeply modified, so the system cannot
have any memory of its configuration at $t_w$. 


As a first partial conclusion we see that the interplay between the 
heterogeneities of the system and the influence of the forcing on different
zones of the sample \cite{bl} leads to two quite different memory effects. 
On one hand the short-time memory is always present and it occurs in a way 
which is counterintuitive with respect to the effect expected
on the basis of the behaviour at constant $x$. On the other hand the 
long-time memory only  exists for a cycle when $x_2 < x_1$ and for 
long times, i.e. for aged systems.

{\bf Perturbation of a stationary state}
The numerical experiments mentioned above have been performed 
during the compaction process, i.e. while the system ages.
Let us now focus instead on the behaviour obtained after a slow decrease 
of the vibration intensity, starting from large forcing: 
this is the usual way to obtain a very compact system \cite{exp-compaction}
which, under weak driving $x_1$, is now in a quasi-stationary state\cite{bl},
i.e. a state where time-translation invariance holds.
This state can now be perturbed by applying a stronger forcing $x_2$ 
for a time interval $\Delta t$, and the
relaxation can be studied when the forcing is set back 
to the original value $x_1$. 
In \cite{joss}, such experiments were performed for small $\Delta t$,
showing  non-exponential relaxations, slower the larger is 
$\Delta t$. We are able here to explore different regimes, varying
$\Delta t$ over four orders of magnitude.

\begin{figure}
\centerline{
       \psfig{figure=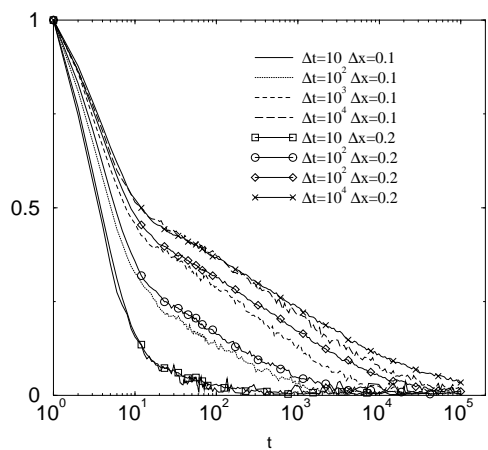,width=6cm,angle=-90}
}
\caption{Relaxation of the potential energy of the system after a 
perturbation $\Delta x=0.1$, $0.2$ during $\Delta t$ of a stationary 
state at $x=0.1$ obtained
after a slow decrease of the vibration intensity, starting from 
large forcing.
}
\label{fig:relax}
\end{figure}

One observes a relaxation occurring in two steps (see figure \ref{fig:relax}).
First one has a rapid decrease (during typically 10 time steps)
corresponding to the relaxation of the topmost layers of the interface;
this can be checked by looking carefully at the density profiles during the
relaxation. Later on a much slower, almost logarithmic, decay occurs 
corresponding then to the re-compaction of layers under the interface.
As $\Delta t$ grows, deeper and deeper layers get involved during the
decompaction process; as $x$ is lowered again, the first relaxation, which
compactifies the interface, makes it more difficult for the lower layers 
to get as compact as before: the relaxation therefore slows down.


Although the very dense medium, if shaken at constant $x$, is in a
stationary state, the perturbation at larger $x$ leads to 
same mechanisms that are also at work during the aging. Therefore,
it is not so surprising to find a slow re-compaction process
after the perturbation.

The precise form of the relaxation can be expected to be model
dependent, at least at short times: for example, in the case
of ``tapping'', the first, rapid part 
of the relaxation does not exist. Therefore, we do not
investigate it in details, and underline only the
most relevant features: even though the system seems in ``equilibrium'' 
(as shown in \cite{bl}, the dense system is stationary and does
not display aging), the relaxation is non-exponential and very slow.

In summary, we have shown that the Random Tetris Model
reproduces the complex phenomenology appearing 
in real experiments for large and abrupt changes of the driving: 
at small times after the change, the reaction of the system is opposite 
to what  could be expected looking at the constant $x$ compaction curves:
if $x$ is decreased, the system compactifies better, while
it gets decompactified for an increase in $x$.
At larger times, the compaction rate crosses over to the one at constant
$x$: this effect was therefore called 'short-term memory'\cite{joss}.

Moreover, we have investigated the return to the previous value
of the driving, in various time regimes, and shown that
an increase in $x$ erases quickly the memory of the previous configurations
while, at long times, a decrease in $x$ can ``quench'' the system
until the previous value of $x$ is restored and the compaction starts again
from a configuration statistically equivalent to the one perturbed at $t=t_w$.

It is noticeable that such a behaviour shares some similarities 
with the phenomena of rejuvenation and memory observed in the context 
of spin glasses \cite{SG}. This feature pushes forward the similarities 
between (non-thermal) granular media and (thermal) glasses and spin-glasses. 
For example, note that a trap model
\cite{bouchaud}, widely exploited for the understanding of the phase space
of spin glasses, has recently been used to describe
granular media \cite{head}, as well as a random-walk model \cite{pouliquen}.
These approaches also reproduce qualitatively the experimental results
of \cite{joss}; however they describe the behaviour of a system in
phase space. It seems to us that the study of
the heterogeneities (and therefore of real space models) could be 
crucial for the understanding of the basic mechanisms underlying 
the dynamics of these systems. In this respect granular media 
seem more tractable than spin glasses and  
detailed experiments about the local rearrangements during compaction
could provide valuable information for the understanding
of the basic mechanisms at work in dense granular media.

%
%
%

\end{document}